# Deep Mediterranean turbulence motions under near-homogeneous conditions

by Hans van Haren

Royal Netherlands Institute for Sea Research (NIOZ), P.O. Box 59, 1790 AB Den Burg, the Netherlands.
e-mail: hans.van.haren@nioz.nl

**Abstract.** Very weakly density-stratified, near-homogeneous 'NH' conditions are found in the deep Western Mediterranean Sea. Under these conditions, over vertical ranges of several hundreds of meters water temperature varies only a few 0.0001°C and the buoyancy frequency is smaller than the local inertial frequency. While such waters are characterized as 'quiescent', they are not stagnant and demonstrate regular bursts of turbulent overturns across scales larger than 10 m that are relevant for deep-sea life. As will be shown from a 3D mooring-array with nearly 3000 high-resolution temperature 'T-'sensors, consecutive NH conditions can last up to a fortnight, before stratified waters are advected over the array. At the site, NH conditions occur about 60% of the time. The majority of NH periods is governed by convection turbulence that is driven by geothermal heating from below. The associated turbulence dissipation rate, which is calculated from Ellison scales after precise band-pass filtering, compares with historic geophysical heat-flux measurements. Convection turbulence leads to buoyancy-driven scaling of spectra, not only of temperature in the turbulence range, but also suggesting extensions across the internal-wave band into sub-mesoscales, and (limited observations of) kinetic energy and waterflow differences. Such spectra are found to be uniform over the 124-m vertical T-sensor range above the flat seafloor. Small spectral deviations are observed when very weakly stratified waters are advected sideways or from above, whereby turbulence levels increase by about 30%. Movies show the alternation between calm periods, turbulent clouds passing, and geothermal-heat flares of various sizes.

# 1 Introduction

The more than 2000-m deep Western Mediterranean Sea is characterized by milli- to centi-degree variations in temperature that demonstrate all dynamics. Above the generally flat seafloor in the vicinity of its northern continental shelf these dynamics include a boundary flow (e.g., Crepon et al., 1982), with meanders and eddies varying at 1-10 days and 1-10 km sub-mesoscales, as well as at 10-30 days and 10-100 km mesoscales. The eddies result from instabilities of the boundary flow and associated fronts, horizontal density variations. These variations are strongest near the surface, but can extend all the way to the seafloor in weaker form. Shorter-period variations involve internal waves, notably having near-inertial periodicities, as transients in response to passages of atmospheric disturbances induced, e.g., by varying winds from the nearby mountain ranges like the Alps blowing over the sea-surface. The near-inertial motions may be trapped by the (sub-)mesoscale eddies (Kunze, 1985), and set-up shorter scale internal waves that eventually dissipate the energy via wave breaking. Although the lack of substantial tides reduces the amount of internal wave energy by about 60% (Wunsch and Ferrari, 2004), the remaining governing processes make the Mediterranean Sea a sample for ocean-dynamics processes (Garrett, 1994).

Detailed observations of these processes and the quantification of the turbulence generated in the deep sea are scarce. Especially also, because at large depths all dynamics occur under weakly stratified conditions requiring high-resolution instrumentation. The mean buoyancy frequency in the deep Western Mediterranean has typical values of $N = O(f)$, f denoting the inertial frequency or vertical Coriolis parameter (van Haren and Millot, 2003), see also the local standard shipborne observations in Fig. 0a, b. Under such weakly stratified conditions internal waves extend well into the sub-inertial sub-mesoscale range, because not only gravity but also Earth's rotational momentum play a role as restoring force (LeBlond and Mysak, 1978): inertio-gravity waves 'IGW'.

Very weak stratification may extend several 100's of meters above the seafloor. In this vertical range a large three-dimensional '3D' mooring-array has been deployed with nearly 3000 sensitive temperature 'T-'sensors measuring half a cubic hectometer for over one year. It was constructed to study development in 3D of turbulence generation, also under weakly stratified conditions. As the conditions push limits to the instrumentation, extensive post-processing is required to capture the physical signals.

In this paper, observations over a 17-day long period of deep-sea turbulence development are presented under near-homogeneous 'NH' conditions. Here, NH is defined as a daily-averaged (pressure-corrected) temperature difference smaller than 0.0002°C (Fig. 0a) between uppermost and lowest T-sensors at nominally h = 125.5 and 1.5 m above the seafloor, respectively. This results in a mean value of about N = 0.5f. The 17 days is the longest consecutive period of NH in a yearlong record as demonstrated by data from a single-line mooring (Fig. 0c). Three more NH periods in the record also last about a fortnight, before stratified waters overcome local conditions. The goals are to gain better statistics, improved physical insight and 3D development of water motions under NH conditions in the deep sea. Anticipated dominant turbulent dynamics processes are general geothermal heating through the seafloor from below that is considered important for deep-sea circulation (Adcroft et al., 2001; Park et al., 2013; Ferron et al., 2017), and IGW possibly combined with sub-mesoscale motions from above.

**2 Several definitions**

The traditional frequency ($\omega$) range for freely propagating internal waves is $f < \omega < N$, for $N \gg f$ (LeBlond and Mysak, 1978). In this range, motions near f are circularly polarized in the horizontal plane. $N^2$ is considered the 'large-scale mean' value of stable vertical density stratification, averaged over at least the inertial period $T_f = 2\pi/f$ and typically 100-m vertical scales to cover turbulent overturns of all sizes. However, observed stratification is not uniformly layered and highly varies with time 't' and space 'x,y,z', being, for example, deformed by the passage of internal waves it supports.

Thus, $N_{min} < N(x,y,z,t) < N_m < N_{max}$, where minimum $N_{min}$ and maximum buoyancy frequency $N_{max}$ are defined for a given area and period in t. Hereby, somewhat arbitrary definitions are used for the vertical scale, which is limited by the separation distance of instrumentation (measuring) rather than by the physical 0.01-0.001-m turbulence dissipation scale. Practically in the present data, smallest and largest vertical scales are 2 and 124 m, respectively. As maximum and minimum values rarely occur, basically only once in a record, their effects on internal-wave band extension are quite small. A more practical extension is the mean $N_m$ of maximum 2-m-small-scale buoyancy frequencies calculated for all vertical profiles. Although NH conditions are characterized by very weak mean stratification, one order of magnitude larger $N_m$ (and thus $N_{max}$) are observed, also under such conditions.

Like mean N being non-existent in the ocean and deep sea for any prolonged period of time at any position, the local vorticity or rotational motion may vary around latitudinal($\varphi$)-dependent planetary value $f = 2\Omega\sin\varphi$ as well. $\Omega$ denotes the Earth rotational frequency. Time- and space-varying waterflow differences such as in eddies can generate relative vorticity of $|\zeta| = f/2$ in the upper-mid-depth of the Western Mediterranean (Testor and Gascard, 2006), of which anticyclonic warm-core eddies can live up to a year. In summer in the deep Ligurian Sea, the 'effective' near-inertial band may widen to $0.9f < f_{eff} < 1.1f$ (van Haren and Millot, 2003). Such frequency modification may add to near-inertial band widening due to latitudinal variation (LeBlond and Mysak, 1978), which however can only lead up to 15% change in f in the Mediterranean due to limited basin-size in meridional direction. Besides a modification of near-inertial frequency (Perkins, 1976), anticyclonic eddies can trap downward-propagating near-inertial waves (Kunze, 1985).

A considerable extension of the internal-wave band under NH conditions is associated with the effect of traditionally-neglected horizontal Coriolis parameter $f_h = 2\Omega\cos\varphi$. For $N = O(f)$, minimum IGW-bound $\omega_{min} \leq f$ and maximum IGW-bound $\omega_{max} \geq 2\Omega$ or N, whatever is largest. The IGW-bounds are functions of N, latitude $\varphi$ and direction of free-wave propagation (LeBlond and Mysak, 1978; Gerkema et al., 2008),

$$\omega_{max}, \omega_{min} = (A \pm (A^2-B^2)^{1/2})^{1/2}/\sqrt{2}, \tag{1}$$

in which $A = N^2 + f^2 + f_s^2$, $B = 2fN$, and $f_s = f_h\sin\alpha$, $\alpha$ the horizontal angle to $\varphi$. For $f_s = 0$ or $N >> 2\Omega$, the traditional bounds [f, N] are retrieved from (1). While NH's $N = 0.5f$ would lead to an impossible wave solution under the traditional approximation, (1) allows free-wave propagation, albeit horizontally for one component (e.g., Gerkema et al., 2008). One of the effects of $f_h$ is turbulent convection in slantwise direction in the z,y-plane, so that apparent stable stratification observed in strictly z-direction may actually reflect homogeneous or unstable conditions in the slanted plane. Local shear, e.g. induced by (sub-)mesoscale eddies, may also guide slantwise convection and add, positively or negatively, to the z-y-plane effects of $f_h$.

Under NH conditions and $\zeta = 0$, $\omega_{max} \approx 2\Omega$, the semidiurnal frequency. The minimum absolute value of IGW-bandwidth $\omega_{max}-\omega_{min}$ is found under conditions of $N = 1f$ (for propagation in meridional direction). The IGW thus widens under more and under less stratified conditions. Under $N = 0.5f$, the bandwidth exceeds half an order of magnitude, and reaches one order of magnitude under $N = 0.22f$. A one-order of magnitude IGW bandwidth is also found approximately under $N = 8f \approx N_{max}$, in the deep Mediterranean. As the deep sea varies its conditions with time and space, the IGW-bounds vary accordingly. This also effects sub-mesoscale motions which are generally thought to be sub-IGW while turbulence is super-IGW. Strict frequency bounds do not exist, so that overlap between the ranges is unavoidable.

## 3 Materials and Methods

The half-cubic-hectometer of deep Mediterranean seawater was measured every 2 s using 2925 self-contained high-resolution NIOZ4 T-sensors, which can also record tilt and compass. Temperature-only sensors were taped at 2-m intervals to 45 vertical lines 125-m tall. Each line was tensioned to 1.3 kN by a single buoy above. Three buoys, evenly distributed over the array, held a single-point Nortek AquaDopp current meter CM that measured wateflow once per 600 s. The lines were attached at 9.5-m horizontal intervals to a steel-cable grid, which was tensioned inside a 70-m diameter steel-tube ring that functioned as a 140-kN anchor. This 'large-ring mooring' was deployed at the <1° flat and 2458-m deep seafloor of 42° 49.50′N, 006° 11.78′E, 10 km south of the steep continental slope in the NW-Mediterranean Sea, in October 2020 (van Haren et al., 2021).

Probably due to a format error, the T-sensors switched off after maximum 20 months of data-recording. As with previous NIOZ4 T-sensors (for details see van Haren, 2018), the individual clocks were synchronised to a single standard clock every 4 hours, so that all T-sensors recorded data within 0.01 s. About 25 T-sensors failed mechanically. After calibration, some 20 extra T-sensors are not further considered due to general electronics (noise) problems. During post-processing and analysis of data under NH conditions specific limitation of the instrumentation was revealed concerning 'short-term bias', which required a removal via low-pass filtering 'lpf' of about 100-300 records, depending on the type of analysis, elaborated below.

Common, primary, NIOZ4 electronic drift or bias of typically 0.001°C mo$^{-1}$ after aging was corrected by referencing daily-averaged vertical profiles, which must be stable from turbulent overturning perspective in a stratified environment, to a smooth polynomial without instabilities (van Haren et al., 2005; van Haren and Gostiaux, 2012). In addition because vertical temperature (density) gradients are small in the deep Mediterranean, reference was made to periods of typically one-hour duration that were homogeneous with temperature variations less than instrumental noise level (van Haren, 2022). Such periods existed on days 350, 453, and 657. This secondary correction allowed for proper calculations of turbulence values using the Thorpe (1977) method of reordering unstable density (temperature) profiles under weakly stratified conditions. Under unstable conditions such as dominated by convection induced via geothermal heating through the seafloor, turbulence values are calculated using the method of Ellison (1957).

Analyzing atmospheric data, Ellison (1957) separated time series of potential temperature θ(t, z), at a fixed vertical position z in two,

$$\theta = \langle\theta\rangle + \theta', \qquad (2)$$

where <.> denotes the lpf series and the prime its high-pass filtered 'fluctuating' equivalent. With multiple sensors deployed in the vertical for establishing a mean vertical temperature gradient, a root-mean-square 'Ellison-'scale $L_E$ can be defined as,

$$L_E = \langle|\theta'_{bpf}|/d\theta_r/dz\rangle, \qquad (3)$$

in which <.> is seen as averaging, the subscript r denotes a reordered, monotonically stable profile, and 'bpf' band-pass filtering in t and z.

Like Thorpe (1977) displacement scales $L_T$, the $L_E = L_O/c$ may be compared with the Ozmidov (1965) scale $L_O$ of largest possible isotropic turbulent overturns in stratified waters, so that the turbulence dissipation rate reads,

$$\varepsilon = c^2 L_E^2 N^3, \qquad (4)$$

in which the constant c needs to be established as a mean from a large distribution of values. If one takes an average value of c = 0.8 (Dillon, 1982) as follows from comparison of $L_T$ with $L_O$, $L_E \approx L_T$ was found under well-stratified conditions above slopes of Mount Josephine, an Atlantic-Ocean seamount (Cimatoribus et al., 2014).

While the Thorpe (1977) method is most sensitive for the largest vertical overturn scales, the Ellison (1957) method is most sensitive for the appropriate separation between internal waves and turbulent motions (Cimatoribus et al., 2014). This separation is not a straightforward task under weakly stratified and NH conditions, like occurring in the deep Western Mediterranean. The method requires precise bpf of turbulent motions from moored T-sensor records, thereby removing instrumental flaws like noise and short-term bias as well as non-turbulent physics signals such as internal waves and sub-mesoscale motions.

Under NH conditions with very low temperature variance, a tertiary correction involved lpf of data. Three phase-preserving double elliptic filters are applied (Parks and Burrus, 1987). One filter is used per record in time with cut-off at 700 cpd (cycles per day). Specific for the Ellison (1957) method, a high-pass filter is used with cut-off at twice the mean $N_m$ of maximum 2-m-small-scale buoyancy frequencies that approximately amounts $2N_m$ = 3-4 cpd. A third filter is used per line in the vertical with cut-off at about 0.05 cpm (cycles per metre). The latter is designed as tertiary correction for removal of short-term bias, which manifests in records shorter than one day under NH conditions in water. This bias is likely associated with weak variable response to temperature changes. Primary corrections are not necessary for spectral internal-wave and turbulence investigations, because that instrumental drift manifests itself as a slow variation with time, typically at a scale of a month, which is well outside daily and shorter time scales. In contrast, short-term bias although measuring only <0.0001°C acts within the time/frequency range of internal waves and turbulence, so that tertiary correction is necessary for spectral analysis as will be demonstrated below.

Lpf data are also used to generate 3D movies, after forcing the data of lowest T-sensors to a common mean value because no drift correction is possible between different lines. Movies are created from a sequence of individual jpeg-format images using freeware video-utility VirtualDub (https://www.virtualdub.org/; last accessed 07 January 2026). The presented movies are accelerated by a factor of 1800 with respect to real-time. Thus, a 0.04 m s$^{-1}$ steady-directed waterflow passes the entire mooring-array in about one movie second.

The large number of T-sensors in a small domain are also expected to improve statistics of at least part of turbulence motions. As temperature-variance spectra are commonly rather featureless for deep-sea observations, with small peaks and no gaps, the focus is on comparison of heavily-smoothed (averaged) spectra with turbulence models that describe a certain frequency ($\omega$) band across which the variance $P(\omega) \propto \omega^p$ varies with slope p. (This is easiest verifiable when spectra are plotted on a log-log scale so that the slope forms a straight line). Two models are of interest, the Kolmogorov (1941) – Obukhov (1949) 'KO' model and the Bolgiano (1959) – Obukhov (1959) 'BO' model.

The KO-model describes the equilibrium inertial subrange of turbulence (Tennekes and Lumley, 1972) with a balance between turbulence kinetic energy 'KE'-production and -dissipation. It describes a forward cascade of energy, with turbulence predominantly produced by vertical current shear in which scalars are passively transported (Warhaft, 2000). As a result, the spectral slopes for KE and a scalar like temperature (variance) are identical in this range and amount p = -5/3.

The BO-model describes the buoyancy subrange of convection turbulence that is actively driven by a scalar. Its direction of energy cascade is still under debate and presumed partially forward and partially backward (Lohse and Xia, 2010). Its spectral slopes are p = -11/5 for KE and p = -7/5 for scalars like temperature variance.

The KO- and BO-turbulence models are compared with various other spectral models, including that of intermittency 'Im' of chaotic processes characterized by p = -1 (Schuster, 1984) and internal wave 'IW' under well-stratified conditions for frequency range f << $\omega$ << N characterized by p = -2 (Garrett and Munk, 1972). The latter slope also more general technically describes spectral slopes of finestructure contamination in time series records of finite length (Phillips, 1971; Reid, 1971). It is found to describe standing-wave induced turbulence in stratified waters of the deep Mediterranean waters and is probably relevant for the open ocean (van Haren et al., 2026).

As for dominant processes governing turbulence dissipation rate in the deep Western Mediterranean, Ferron et al. (2017) calculated three-times larger values due to geothermal heating 'GH' that due to IGW breaking. Their calculations were based on sparse shipborne microstructure profiling. Spectral analysis may distinguish between the two when leading to convection or shear turbulence. For reference of periods with dominant GH, the mean local geophysics-determined GH amounts 0.11 W m$^{-2}$ (e.g., Pasquale et al., 1996). After conversion of the heat flux into buoyancy flux 'Bfl' in the overlying waters, GH-induced turbulence dissipation rate is calculated as,

$\varepsilon_{GH}$ = Bfl/$\Gamma_C$ = 1.2×10$^{-10}$ m$^2$ s$^{-3}$. (5)

In previous data (van Haren, 2025) the mixing coefficient was found to amount $\Gamma_C$ = 0.5, which is typical for convection turbulence (Dalziel et al., 2008; Ng et al., 2016).

## 4 Results

Instrumental drift is not well visible from the two temperature records in the NH data-overview time series (Fig. 1a), although variations of 0.0001°C are visible in a plotted range of 0.0012°C. (Here and elsewhere, 'temperature' is used in short for pressure-corrected Conservative Temperature (IOC et al., 2010). Long-term drift has been simply 'removed' by de-trending the two records before plotting. While 2-3-day periodic variations have excursions of about 0.0005°C, <0.5-day periodic variations generally stay within 0.0001°C excursions. Between the two records obtained vertically 124-m apart, negligible <0.0001°C differences are seen during the first week of observations, while >0.0001°C --positive and negative-- differences occur during the second week. The transition between the two distinctively different vertical temperature differences is observed near day 354.5. This major distinction in vertical temperature difference under NH conditions is compared with several other variables.

Waterflow speed (Fig. 1b) shows a quasi-inertial (0.73-day) periodicity during the first week when small vertical temperature differences are observed, and quasi-constant, longer-periodicity values during the second week when alternating (weakly) stratified and unstable conditions are observed. No obvious distinction is observed in acoustic amplitude (Fig. 1c), which is invariably low under NH conditions. Nearby island atmospheric wind variance (Fig. 1d) shows a 2-3-day periodicity throughout, but with distinctive increase in values after day 357. Employing a 5.5-day advance, it is seen that wind variance governs the overall 2-3-day variation in deep-sea temperature, rather than distinction in vertical temperature difference (of which the second week then associates mostly with smallest wind variance). Daily (and 124-m vertically) averaged turbulence dissipation rate (Fig. 1e) shows hardly any variation from mean GH-value (5) during the first week, and about 50% increase in mean value with 2-4-day periodicity during the second week. The overall mean turbulence dissipation rate for the 17-day period amounts $1.6\pm0.8\times10^{-10}$ m$^2$ s$^{-3}$, which is about one-third larger than mean GH-value (5) and commensurate with the findings of Ferron et al. (2017), if the excess value over (5) is attributable to simultaneous IGW —i.e., standing interfacial-wave breaking rather than ten times more energetic slantwise internal-wave convection (van Haren, 2026a submitted). The excess value compares with open-ocean low turbulence dissipation rates well away from boundaries (e.g., Yasuda et al., 2021). In the Mediterranean tides are so small that direct internal wave sources are reduced by 60% compared with the open ocean.

**4.1 A six-day movie of warmer and cooler water 'columns'**

The six-day single-line data overview from the first week of NH in Fig. 1 shows vertically uniform temperature values, which vary ±0.0001°C in time (Fig. 2). The 3D movie shows that the vertical-uniformity exists less than half the depicted period. During the remainder, either heating is seen from below (e.g., around day 350.5) or from above (day 350.9). Most dramatic are clouds of relatively warm or cool water that become advected through the instrumented array, e.g. from the northeast (day 348) and from the west (days 352-353). Below, a daily narrative describes impressions of the movie.

The mean flow is directed NE at 0.02 m s$^{-1}$, so that the large-ring mooring-array is passed in 0.04 day, which equals to about 2 s in movie-time. The movie starts with uniform warm waters that slowly cool, with relatively warm waters entering sideways from the northeast. The flickering of different colours represents turbulent motions when it extends over several T-sensors. On day 349, warm turbulent clouds are advected from the West, besides warmer waters coming from above. Day 350 shows large GH from below initially, followed by (advection of) warmer waters from the West and ending with warm water coming in from the north near the top. Day 351 demonstrates suppression of GH that remain in the lower h = 10 m from seafloor, with clouds of warmer water advected higher up. Day 352 continues advection of relatively warm cloud passages, coming from the west. Day 353 shows an acceleration of advection-type of the previous day, also down to the seafloor. The movie ends as it started, with near-uniform warm waters from above, even under NH conditions.

Although clearly visible at times, no dominance or persistence in time is observed of GH, which is regularly seen suppressed by motions from above. Nevertheless, turbulence values are close to mean GH-value (5), and possibly advected clouds of turbulent warmer waters are generated by convection 'upflow' elsewhere. For comparison, the second week of data shows slightly more vertical and time variability in temperature (Appendix A).

Resuming, advection seems more important than anticipated as warming is not exclusively found from below or above. Thus, precise sources are unknown, and advected waters can be heated from below or above elsewhere. The amount of turbulence varies strongly over short periods of typically one hour in time, also in the deep sea, which is seen not to be stagnant, also under NH conditions.

**4.2 Short-period details from a single line**

During 2.5 days of dominant near-inertial variation in waterflow speed (Fig. 1b), observed temperature is generally slightly unstable, but otherwise also shows near-inertial period variation with time (Fig. 3a). The variation of temperature is between inertial and semidiurnal periodicity. The vertically averaged turbulence dissipation rate varies over shorter timescales (Fig. 3b), around its average which is equal to (5) well within error. The quasi-uniformity of temperature in the vertical is reflected in hardly any variation in temperature variance spectra (Fig. 3c), except for that of the lowest T-sensor record. In the 31-records smoothed mean for the lower half of T-sensors, slightly more variance is found than for the upper half, with significantly larger values at the lowest T-sensor. Between frequencies $f \leq \omega \leq 500$ cpd, spectra are featureless and adopt the BO-scaling along (unscaled) spectral slope $p = -7/5$, except for the lowest T-sensor which roughly follows Im-slope $p = -1$. Thus, apart from the lower $h = 2$ m above seafloor, motions are dominated by convection turbulence over at least 124-m above seafloor, and across all resolved frequencies including into the IGW. No KO-scaling of inertial subrange is resolved, albeit that the significant temperature variations are close to noise and bias-affected levels.

During a one-day period of dominant but very weak stratification (Fig. 4), several differences are observed with respect to Fig. 3. Stratification is very weak but non-negligible, and some general convection turbulence occurs in the middle half. About 0.1-day periodicity is seen in warm waters from above (Fig. 4a), noting that the movie of Fig. A1 demonstrated that in the first half of this period the warm clouds were merely advected horizontally. The warm waters around day 355.7 came from above. Small-scale GH is seen near the seafloor on days 355.0 and 355.2. The vertically averaged turbulence dissipation rate (Fig. 4b) peaks when relatively warm waters pass, e.g., on days 354.93, 355.1, 355.25 and 355.7. These peaks all extend above GH-value (5). Although the shortness of record does not resolve the IGW band, one-day mean spectra (Fig. 4c) are even more uniform than in Fig. 3, with exception of lowest T-sensor which roughly slopes with Im's $p = -1$ between about $100 \leq \omega \leq 500$ cpd and BO-scaling $p = -7/5$ for $10 \leq \omega \leq 100$ cpd. The spectra at other heights above seafloor have different spectral slope ranges. Between about $60 < \omega < 600$ cpd they slope like $p = -7/5$. At lower frequencies between $6 \leq \omega \leq 60$ cpd, they predominantly slope like $p = -5/3$ although some portion suggests a steeper slope closer to $p = -2$-scaling. They associate with very thin ($\Delta z < 2$ m) short-lived stratified layers.

A half-day example of dominant large GH-flares is shown in Fig. 5. Even in the unstable flares with warming from below, apparently-stable stratified 'layers' occur. While the mean turbulence dissipation rate is 25% above GH-value (5), which is still within error, individual peaks are observed during such apparent stratification such as around days 362.28 and 362.6. The limited spectral range shows slopes between that of Figs 3 and 4, with a tendency to dominant BO-scaling of $p = -7/5$ and significantly larger temperature variance in the lower half of the sensor-range. As before, the lowest T-sensor demonstrates a deviating spectral slope, here between $100 < \omega < 600$ cpd, and which may in part reflect improper application of the vertical lpf. Physically however, the lowest T-sensor is affected by intermittent GH-motions that are found dominant in the lowest $h < 1$ m from the seafloor (van Haren, 2023).

**4.3 Increased spectral smoothing**

Although the spectra of Figs 3c-5c show reasonably uniform slopes, they were calculated in smoothed form for a single-line mooring, without taking advantage of the multiple lines of the large-ring mooring. Improved statistics for records from the 17-day NH period is investigated in this sub-section.

Applying spectral smoothing over data, if not interpolated, from vertically three T-sensors and all 45 lines for two one-week periods demonstrates the variability in random statistics as a function of frequency range (Fig. 6). On average 130 out of possible 135 records were included in the smoothing. The spectra, here presented in unscaled form, show a twice-larger vertical spread (on logarithmic scale) between about $0.6 < \omega < 60$ cpd compared to the rest of the spectrum. This suggests that the internal-wave and large-scale turbulence ranges are variable with time, that they not completely resolved by the array's horizontal scales $O(10)$ m, and that signals are not isotropic incoherent at these frequencies. Nevertheless, the spectra support some findings of those in Figs 3c-5c.

In Fig. 6a, the tendency is a dominant BO-slope of $p = -7/5$, which is found most continuously at heights near the seafloor where largest temperature variance is at super-buoyancy frequencies $\omega > N_{max}$. Largest deviation from $p = -7/5$ slope is around mean small-scale buoyancy frequency $N_m$, adjacent to a small peak around $2\Omega$. The BO-scaling is resumed at sub-inertial frequencies, in the overlap with sub-mesoscale motions. In this range, KE tends to $p = -11/5 > -5/3$, although poorly resolved, but obviously steeper than horizontal waterflow differences dU that tend to $p = -7/5$. This dU-slope suggests that vorticity and/or flow divergence associate with scalar-convection activity. The instrumental noise was too high to resolve KE and dU at super-buoyancy motions. This first week of NH period was characterized by a spectral peak of near-inertial motions, in contrast with the second week.

In Fig. 6b, dU shows hardly significant signals above noise, while KE indicates a small peak at f and a similar steep $p \approx -11/5$-slope around minimum IGW-bound $\omega_{min}$ as in Fig. 6a. Compared with the first week, the temperature spectra in Fig. 6b show slightly more variance, notably at larger heights from the seafloor across $\omega_{min} < \omega < 600$ cpd, while being reduced at sub-mesoscale frequency (hardly resolved). The band-widening is extended to higher frequencies and more at larger heights, which suggests more coherent or less random motions. At all heights above seafloor a small peak is observed around $\omega_{min}$, which reflects the 2-3 day periodicity seen in time series. Because of the increased band-widening, spectral slopes are more difficult to determine, which lie between BO- and KO-scaling.

Increased smoothing is investigated in Fig. 7. Taking advantage of only small variations across the 124-m vertical range during the NH period, smoothing is applied over 17 days and approximately 2000 independent T-sensor (pairs). While temperature is found significantly coherent over 2-m vertical scales at sub-inertial frequencies, it takes two orders of magnitude in frequency to become significantly incoherent at $\omega > 200$ cpd (Fig. 7a). This is reflected in phase difference, which equals zero at most frequencies (Fig. 7b), and in range reduction with respect to random noise in temperature variance (Fig. 7c). Coherence equals about 0.5 at overall maximum 2-m-scale buoyancy frequency $N_{max}$, which suggests that stratified turbulent motions can have some coherence and/or that freely propagating internal waves may exists supported by stratification at finer vertical scales $\Delta z < 2$ m. At about $\omega > N_{max}$, the coherence and associated temperature-variance spectra are affected by short-term bias, as vertical lpf could not be applied on these data. (The cut-off was on ten times larger scales than the 2-m distance between T-sensors so that 2-m coherence would be affected). The deviation between filtered and non-filtered spectra amounts more than half an order of magnitude in the noise range (Fig. 7c). It demonstrates the necessity of application of vertical lpf under NH conditions for other analyses.

As suggested in Figs 3c, 5c, 6a, the smoothed temperature scalar spectrum from the 17-day NH period tends to BO-scaling of $p = -7/5$, which is observed to be significantly tight around $\omega = 100$ cpd, and with larger power-variation extending across IGW to sub-mesoscale sub-inertial frequencies. Between about 0.5 cpd and $N_{max}$, the spectrum is elevated by about half an order of magnitude, as far as can be inferred from the poorer effect of smoothing. The transition between the two levels of BO-scaling lies between approximately $N_{max} < \omega < (2-3)N_{max}$, and has a slope of about $p = -2$.

## 5 Discussion and Conclusions

Technically, the frequency-dependence in effect of spectral smoothing demonstrates the complex statistics of oceanographic data. Only at $200 < \omega < 600$ cpd (transfer to dominant instrumental noise) the smoothing acts like on quasi-random signals, at vertical scales $< 2$ m and horizontal scales $< 9.5$ m. While band-smoothing, averaging of neighbouring spectral bands, reduces the vertical power-variation (cf., Fig. 7c), the effect of poorly resolved statistics extends across about $\omega_{min} < \omega < N_{max}$. The temperature-variance width is largest near $N_{max}$ and seems to reduce again at sub-inertial frequencies.

This relates non-random statistics to largest turbulent overturn, but especially also small internal-wave scales, also under NH conditions in the deep sea.

Overall, convection turbulence is found to be dominant in the lower 125 m above the flat seafloor at the site of the large-ring mooring. While weak stable and unstable portions of 10-30 m scale deviate from vertically pure homogeneity at times, temperature variations are larger with time. The impression from 1D depth-time images is a sequence of alternating relatively warmer and cooler >125-m 'columns'. In contrast, quasi-3D movies show rather turbulent variations within such columns, either driven from below, or from above. Mostly, turbulent clouds are advected sideways pointing at sources above or below upflow.

While the main source of convection turbulence seems clear under NH coditions, being general GH through the seafloor, the observation of its response in the deep-sea waters varies over short spatial and time scales. This may be partially due to suppression by warmer, stratified waters from above, and partially due to geological variability. Flares are commonly a response of nonlinear convection cells in an enclosed basin that is uniformly heated from below (Dalziel et al., 2008; Ng et al., 2016), except perhaps at h < 1 m as observed at our lowest T-sensors.

The additional source leading to nonlinearity and turbulence generation under very weakly stratified NH conditions, presumably via internal-wave breaking from above is less clear. The KE spectra suggest a source at the inertial frequency, but no (peak) response is observed in temperature records at f. Temperature shows two minor peaks, either around semidiurnal frequencies or around about 0.4 cpd which is close to the non-traditional lower IGW bound, at sub-mesoscales.

Whatever the source, in all cases the high-frequency $\omega > N_{max}$ turbulence spectra scale with the BO-model for the buoyancy subrange of stratified turbulence. When dominated by GH from below, the spectra continue to the IGW, and perhaps to 2D sub-inertial eddies, without change in variance level. This could suggest a uniform cascade of temperature variance, although laboratory experiments are inconclusive on turbulence from such a source (Lohse and Xia, 2010). When turbulence appears from above, IGW shows increased temperature variance-variation over the BO-scaled turbulence range, even under NH conditions of very weak stratification.

These observations have also been made during other periods of NH such as between days 310-324, 451-463 and 619-635. Some of these included elevated inertial motions, between days 315-322 and 453-460, all showing maximum one-week duration. In the investigated 17-day record, the first week of elevated inertial motions shows dominant GH with lowest mean turbulence values. During the second week of the NH period, the about 30% increase in mean turbulence values are attributed to weak internal-wave breaking. These turbulence values are close to those observed by Ferron et al. (2017).

Compared with the weak internal-wave breaking under NH conditions, ten times larger mean turbulence dissipation rates occur during strong internal-wave breaking events at the site of the large-ring mooring, in the deep Western Mediterranean where tides are weak. Under such strong-turbulence but stratified-water 'SW' conditions $N \approx (1-2)f$ and the IGW is barely wider than under NH conditions, albeit with a shift in bound-frequencies. SW episodes are investigated elsewhere (van Haren, 2026 submitted).

*Data availability.* Only raw data are stored from the T-sensor mooring-array. Analyses proceed via extensive post-processing, including manual checks, which are adapted to the specific analysis task. Because of the complex processing the raw data are not made publicly accessible. Current meter and CTD data are available from van Haren (2025): "Large-ring mooring current meter and CTD data", Mendeley Data, V1, https://doi.org/10.17632/f8kfwcvtdn.1. Movies to Figs 2 and A1 can be found in van Haren, Hans (2026), "Movies to: Deep Mediterranean turbulence motions under near-homogeneous conditions", Mendeley Data, V1, https://doi.org/10.17632/kw2k6v82f9.1. Atmospheric data are retrieved from https://content.meteoblue.com/en/business-solutions/weather-apis/dataset-api.

*Competing interests.* The author has no competing interests.

*Acknowledgments.* This research was supported in part by NWO, the Netherlands organization for the advancement of science. Captains and crews of R/V Pelagia are thanked for the very pleasant cooperation. NIOZ colleagues notably from the NMF department are especially thanked for their indispensable contributions during the long preparatory and construction phases to make this unique sea-operation successful. I am indebted to colleagues in the KM3NeT Collaboration and L. Gostiaux, who demonstrated the handling of large data sets.

**Appendix A Movie two**

The mean waterflow at h = 126 m for Fig. 1's second-week six-day NH period in Fig. A1 is directed ENE at 0.04 m s$^{-1}$, so that the large-ring mooring-array is passed in 0.02 day, i.e. in 1 s movie-time, half the one in Fig. 2. The movie's daily narrative is as follows. The movie starts, and ends, cooler than the one of Fig. 2. On day 355, warm waters are advected from the west near the top, initially accompanied by suppressed GH near the seafloor. Towards the end of the day a wavy-like warming comes in from above with frequent un- and down-going quasi-standing motions. The cooler day 356 starts with GH and ends with a split of rapid up-down stratified motions near the top and GH near the seafloor, with intense short-range variations between lines. On day 357 GH continues, but gradually the warmer waters become more uniform and seem to be advected from above. During the first half of day 358, clouds of variable temperature move mostly horizontally. GH is dominant during the second half. On day 359, the uniform warm waters seem to be initiated via GH, although later warm clouds also come from above and sideways, via relatively calm advection. On day 360, the advection becomes more intense, with warm-water clouds from above before cooler waters are advected in. With respect to Fig. 2, the movie in Fig. A1 has a 60% larger temperature range and shows more variation in episodes of advection sideways as well as from above.

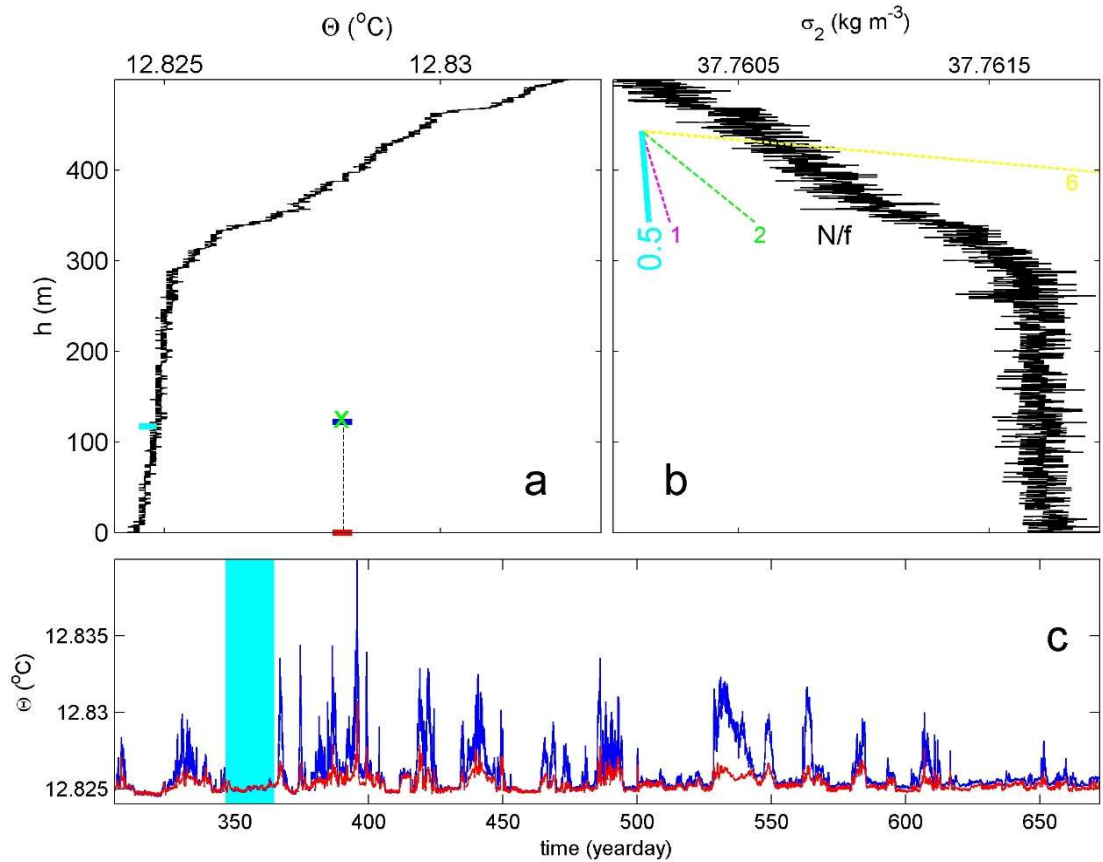

**Figure 0.** Common oceanographic data overview from single shipborne Conductivity Temperature Depth 'CTD' profile and single mooring line. (a) Conservative Temperature (IOC et al., 2010). Lower 500 m of CTD-profile, with indications for height above seafloor of moored instruments: current meter (x), uppermost (blue) and lowest (red) temperature sensor. The small cyan bar indicates the temperature range across the vertical extent of the moored instrumentation. (b) Corresponding density anomaly referenced to a pressure level of $2\times10^7$ N m$^{-2}$. Several vertical density-stratification rates are indicated in terms of the ratio N/f of buoyancy over inertial frequency. The cyan line indicates the ratio corresponding to that of the vertical mooring extent in the CTD-profile. (c) Yearlong time series of detrended Conservative Temperature from uppermost and lowest T-sensors of mooring line 24. The highlight indicates the 17-day period under near-homogeneous 'NH' conditions discussed in this paper. Time is in days of 2020, +366 in 2021.

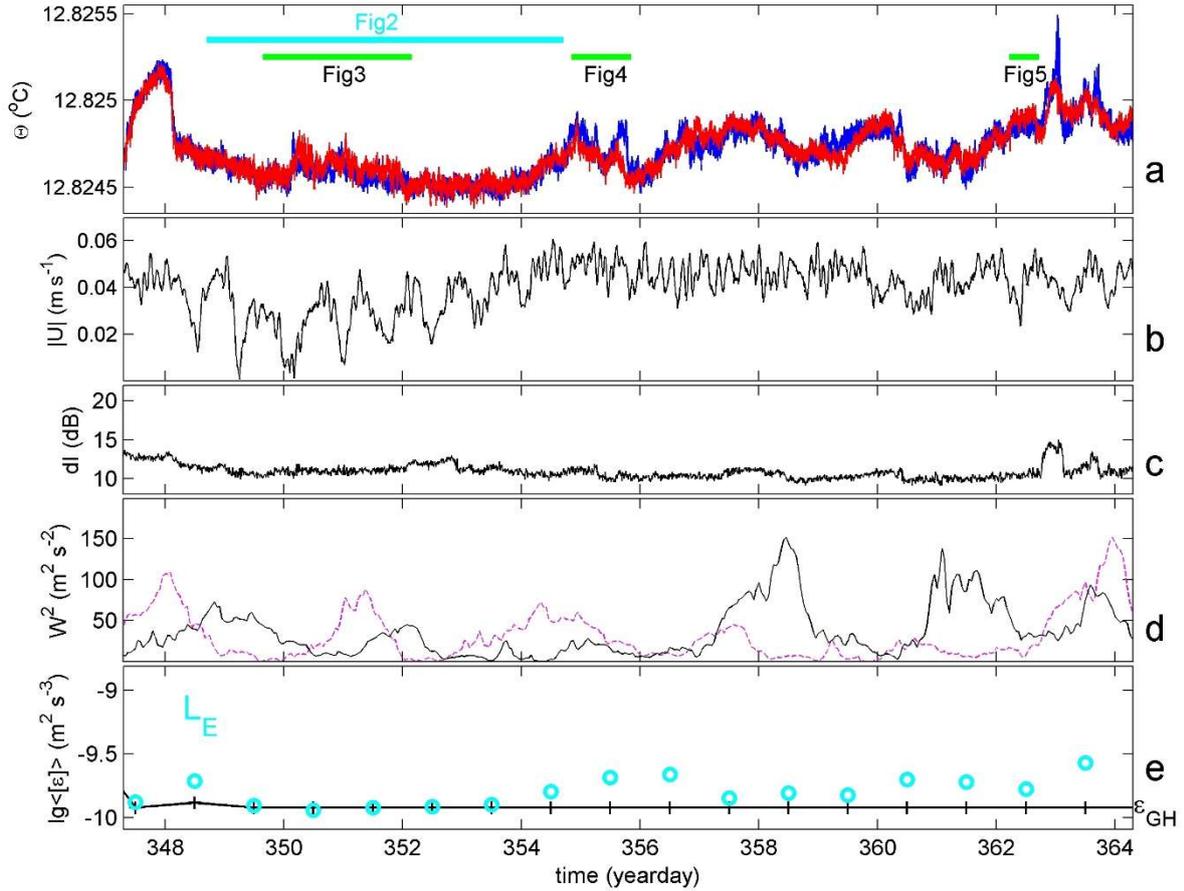

**Figure 1.** Fortnight, 17-day long period of NH conditions at the large-ring mooring. Time-series overview of several environmental parameters. (a) Conservative Temperature from line 24, T-sensors at h = 0.7 (red) and 124.7 m (blue) above seafloor. Unfiltered, bias-corrected and 10-s sub-sampled data. Horizontal colour bars indicate the time periods of Figs 2-5. (b) Waterflow amplitude measured at h = 126 m above seafloor, hourly low-pass filtered 'lpf'. (c) As b., but for relative acoustic-echo amplitude. (d) Wind variance measured at island-station 'Porquerolles', about 20 km north of the mooring. In magenta, 5.5 days shifted forward. (e) Vertically and daily averaged turbulence dissipation rate from Ellison (1957) scales '$L_E$' (o), and values (+) fixed to nominally geothermal heat 'GH' dissipation rate of $1.2\times10^{-10}$ $m^2$ $s^{-3}$ (van Haren, 2026b submitted) when daily-mean $\Delta\Theta < 0.0002$°C over $\Delta z$ = 124 m.

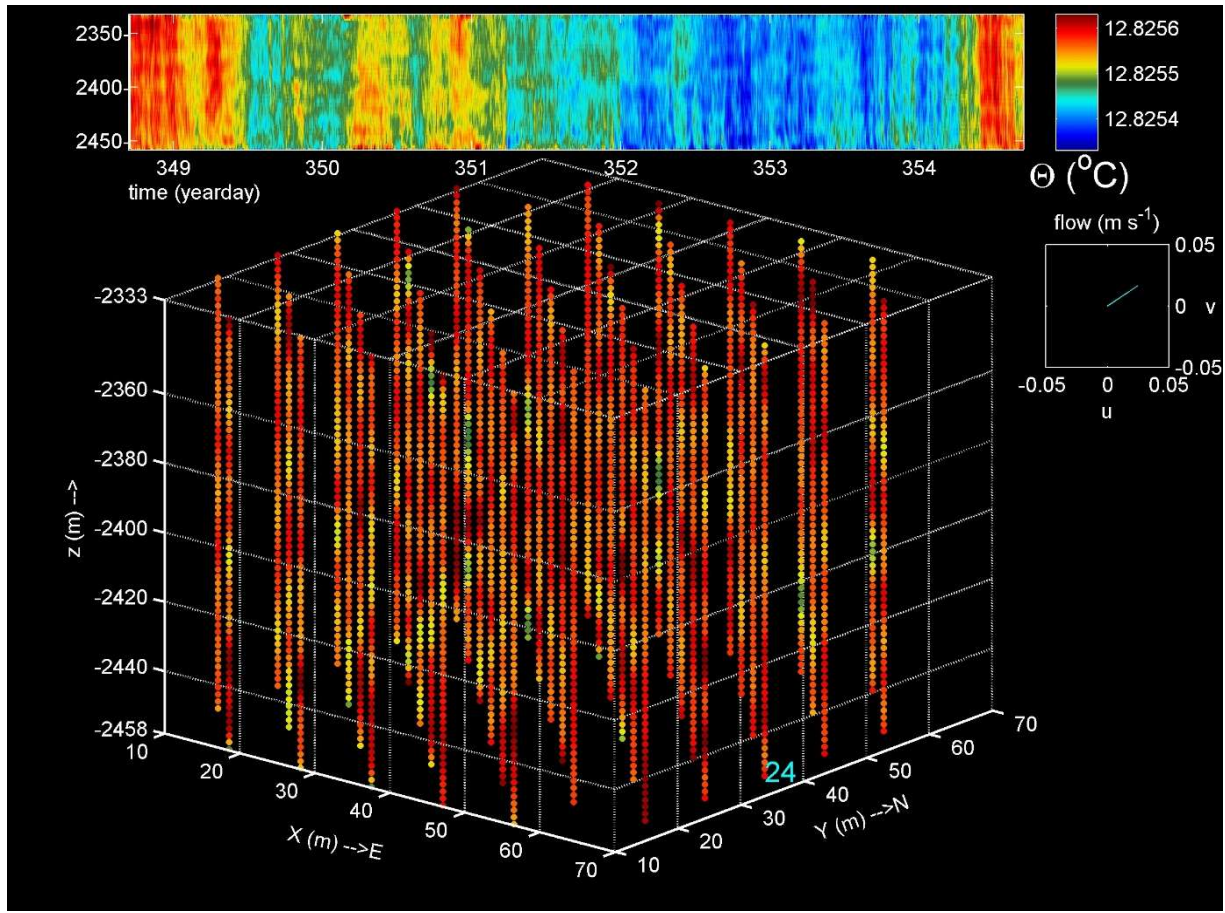

**Figure 2.** Quasi-3D movie from six days of lpf temperature data from about 2800 T-sensors in nearly 0.5-hm$^3$ mooring-array. Each sensor is represented by a small filled circle, of which the colour represents Conservative Temperature in the scale above (total range: 0.0003°C). In the movie, above the cube, which is vertically depressed by a factor of two, a white time-line progresses in a 6-d/124-m time/depth image of temperature data from line 24 on the east-side of the cube. The 288-s movie is accelerated by a factor of 1800 with respect to real-time. In the small panel to the right in this figure, but not appearing in the movie, the mean flow is indicated, measured at h = 126 m above seafloor.

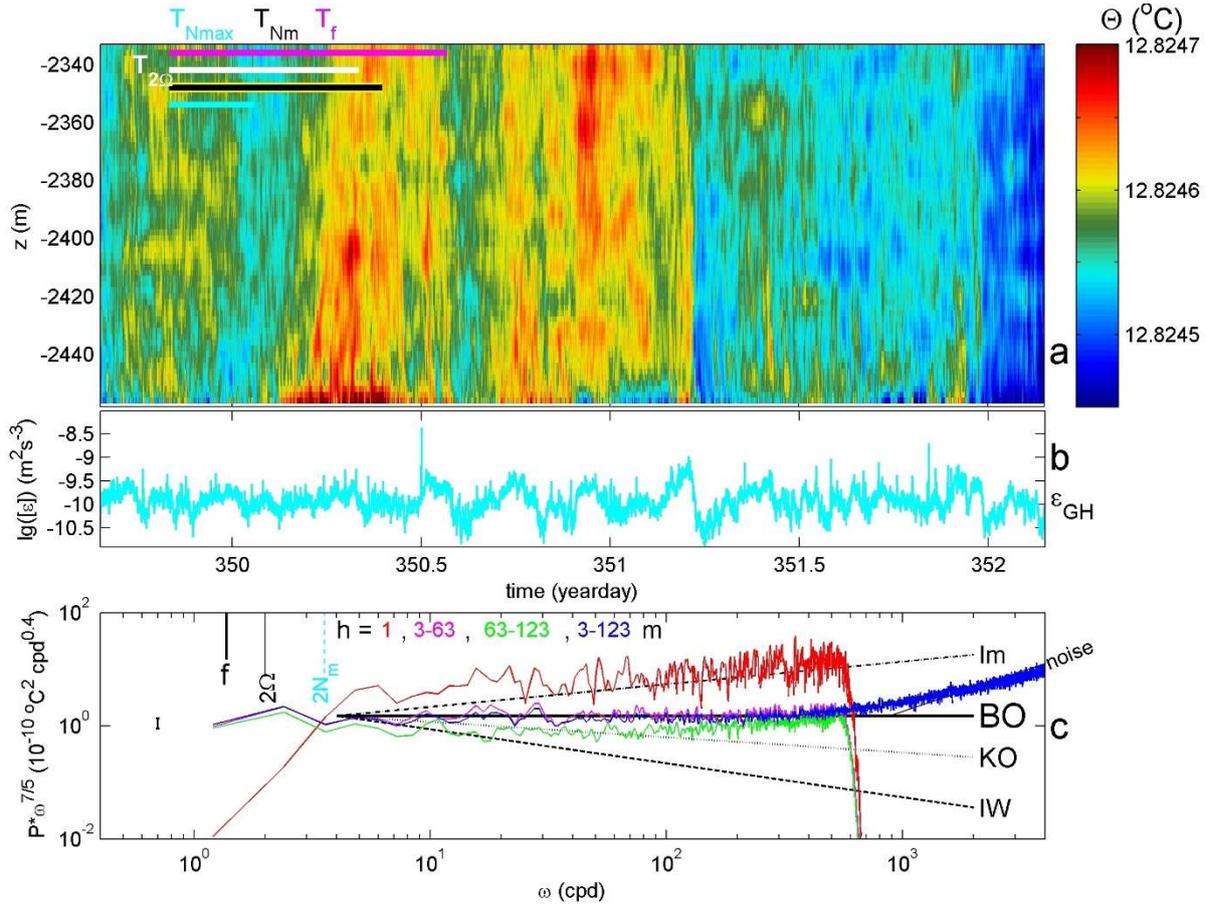

**Figure 3.** Detail of 2.5-day of moored T-sensor data from corner-line 47 during the inertial-motions dominated part of Fig. 1. (a) Time/depth series of 700-cpd and 0.05-cpm lpf Conservative Temperature after post-processing, as outlined in Section 2. Horizontal bars indicate one inertial period $T_f$, one semidiurnal period $T_{2\Omega}$, one mean of maximum 2-m small-scale buoyancy periods $T_{Nm}$, and one maximum small-scale buoyancy period $T_{Nmax}$. The mean 124-m large-scale buoyancy period equals about $T_N \approx 2T_f$. Over the entire colour range, the temperature difference amounts $\Delta\Theta = 0.00025°C$. (b) Time series of 124-m vertically averaged turbulence dissipation rate established using Ellison (1957) scales with cut-off frequencies at $2.2N_m \approx 0.9N_{max} \approx 4$ cpd and 700 cpd. (c) Weakly smoothed (about 10 degrees of freedom 'dof') spectrum of [4, 700] cpd bpf data of lowest T-sensor (red) in comparison with 700-cpd lpf heavily smoothed (300 dof) lower (magenta) and upper (green) layer spectra, and with 900 dof (cf. error bar) 124-m averaged non-time-filtered spectra (blue). All spectra are for time series after 0.05-cpm vertical lpf to remove short-term bias. They are scaled with buoyancy-driven model $\omega^p$, BO-slope $p = -7/5$, for scalar quantities (see text). Several spectral slopes are indicated by straight lines, with their exponent values for unscaled spectra. Given are inertial frequency f, semidiurnal $2\Omega$, and $2N_m$.

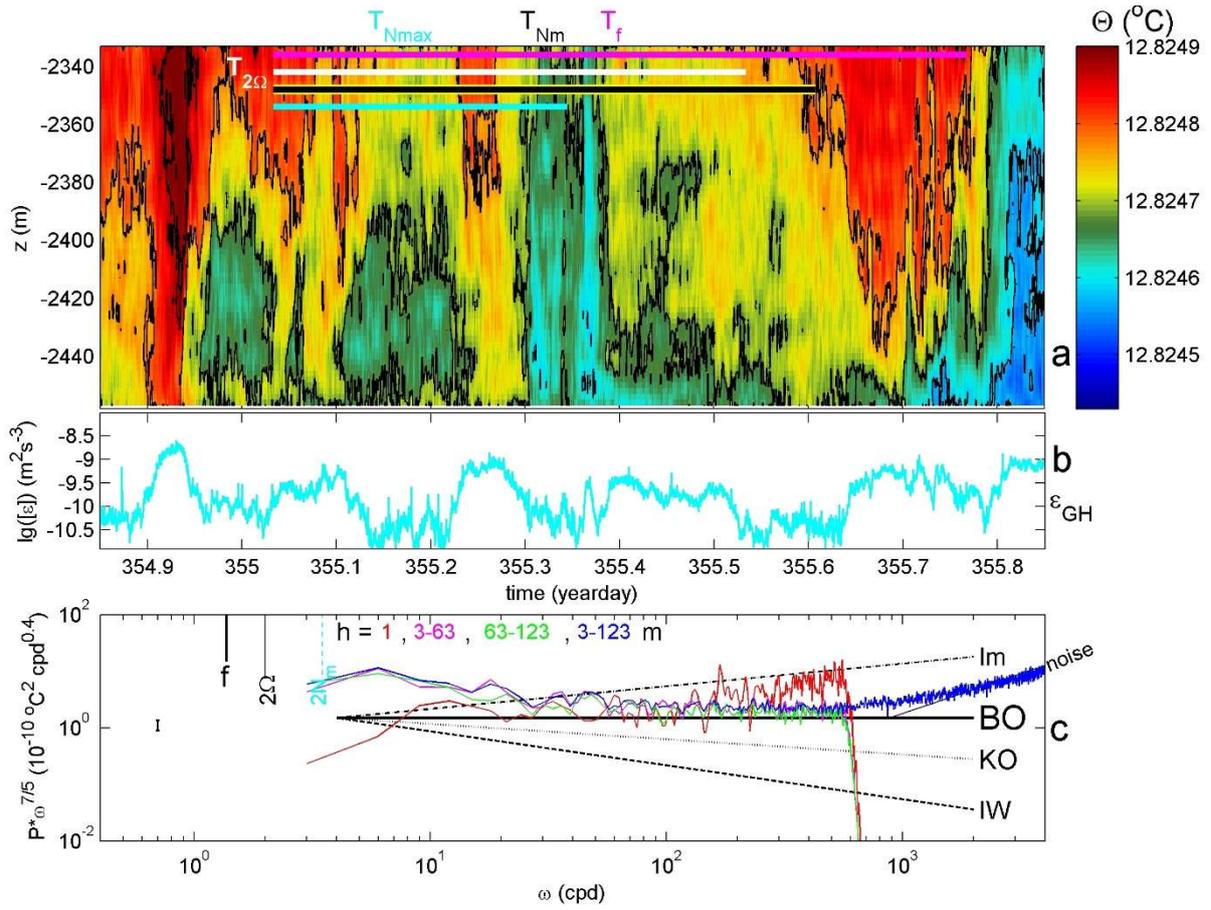

**Figure 4.** As Fig. 3, but for 1.0-day example of spectral uniformity between upper and lower T-sensors under very weak but stable stratification, except for some intermittency-slope at the lowest sensor. In a., the temperature difference amounts $\Delta\Theta = 0.00047°C$ over the entire colour range; black contours are drawn every $0.0001°C$. In c., $2N_m \approx 1.1N_{max}$.

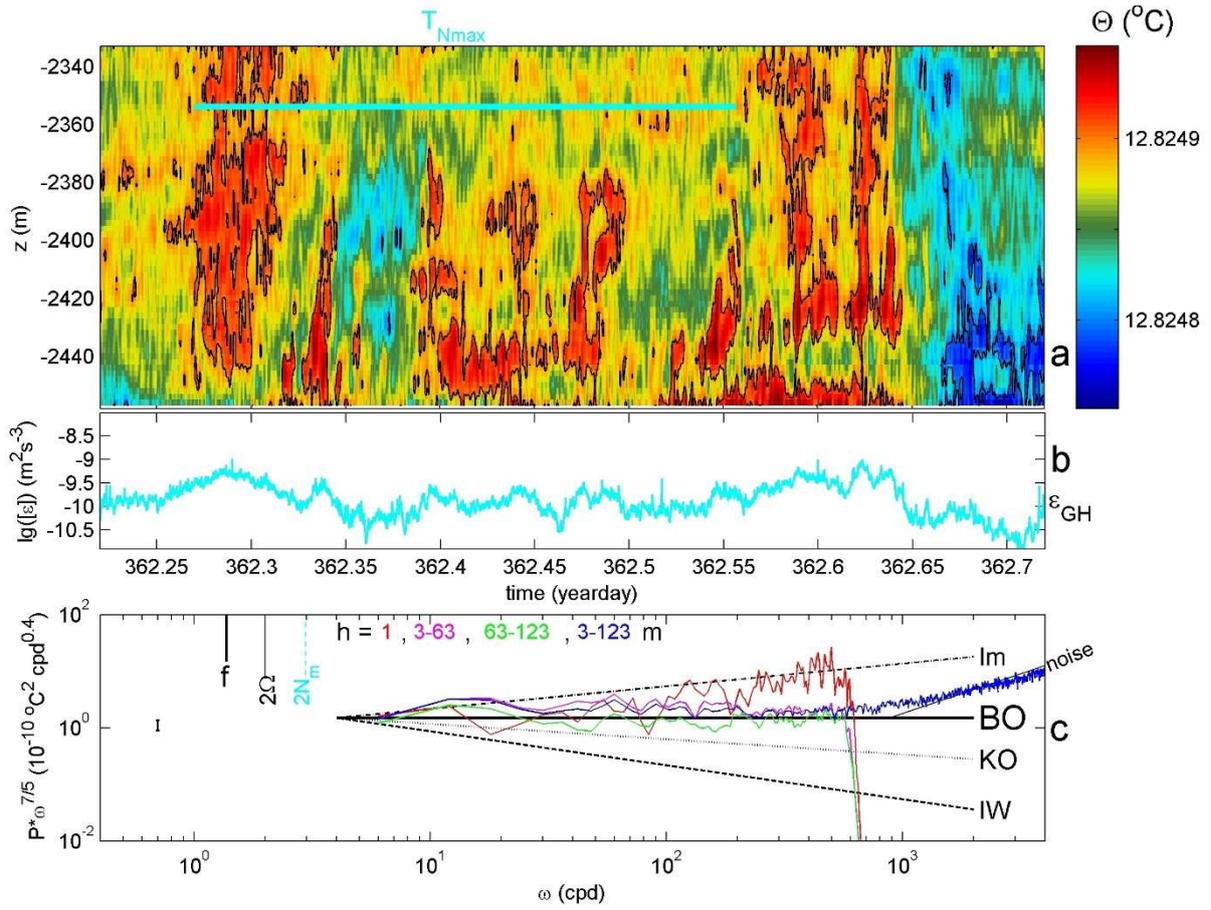

**Figure 5.** As Fig. 4, but for 0.5-day example under partially unstable conditions. Over the entire colour range, the temperature difference amounts $\Delta\Theta = 0.0002°C$. In c., $2N_m \approx 0.9N_{max}$.

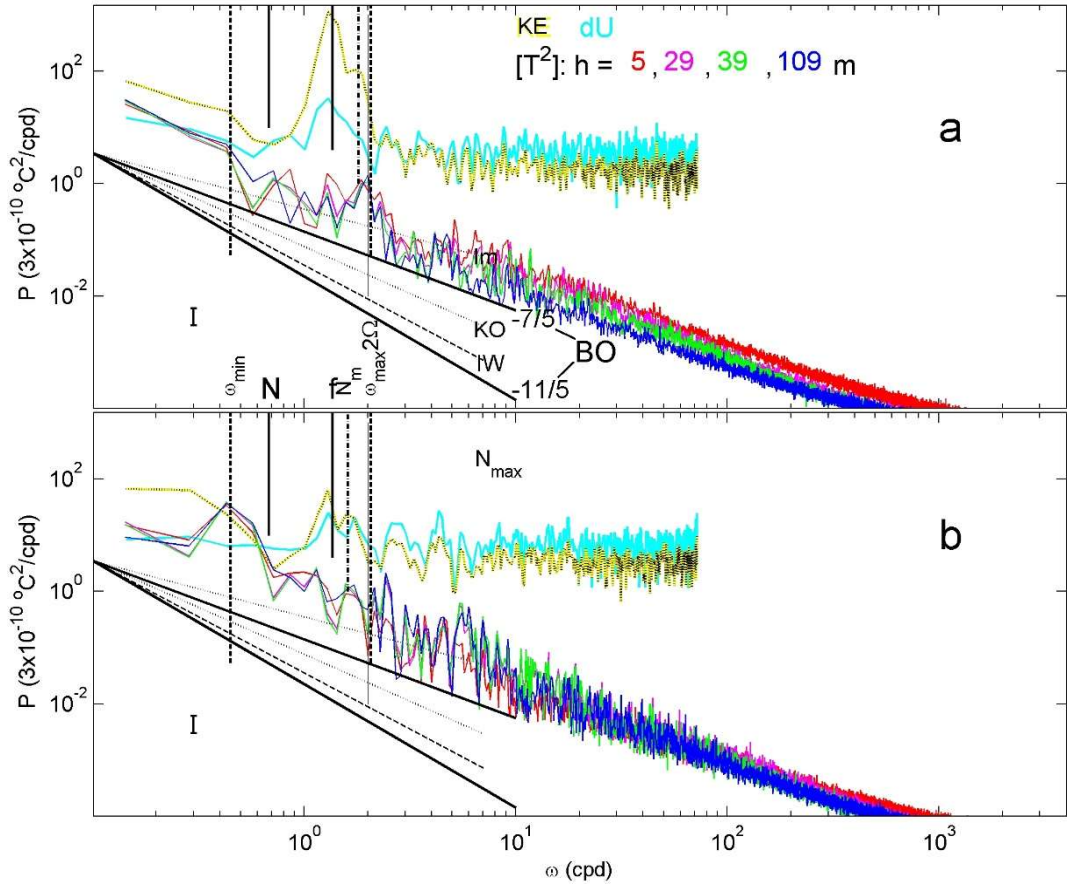

**Figure 6.** Unscaled spectra for two non-overlapping one-week NH periods of Fig. 1. Kinetic energy 'KE' and horizontal waterflow differences 'dU' averaged over the three current meters at h = 126 m (arbitrary vertical scale) are compared with non-filtered temperature variance 'T$^2$' averaged over all 45 lines and over three, if not interpolated, vertically neighbouring T-sensors around heights above seafloor as indicated. For reference, model spectral-slopes and some frequencies are given, see text. (a) Days 347.3-354.3. (b) Days 355.3-362.3.

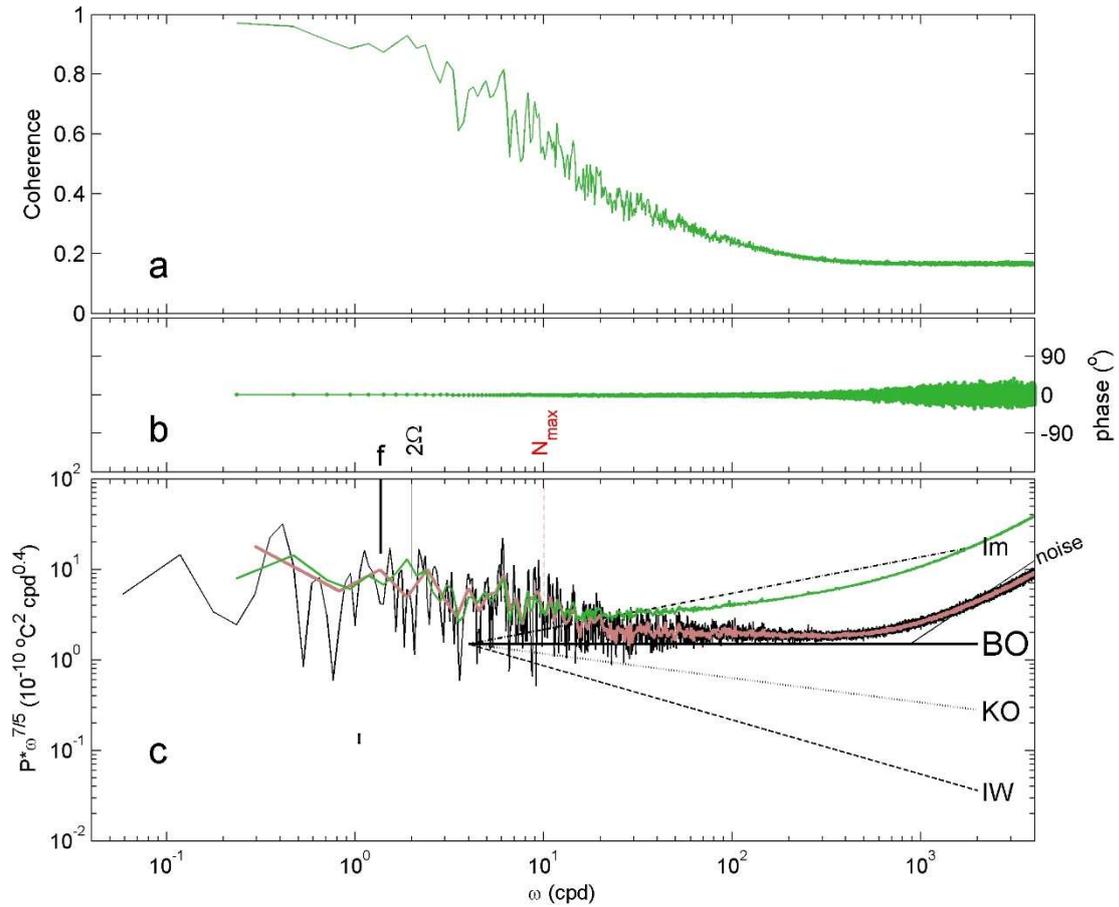

**Figure 7.** Spectral statistics for 17-day period of Fig. 1 for all independent, not-interpolated moored T-sensor data, excluding those from uppermost and lowest T-sensors. (a) Heavily smoothed (~7000 dof) coherence over 2-m vertical distances of 1957 T-sensor pairs without vertical lpf application. (b) Corresponding phase. (c) Corresponding temperature variance (green), which is compared with average over 2088 T-sensors for which the vertical lpf is applied (black; 10 times more band-smoothed in pink). Spectra are scaled with scalar-BO $\omega^{-7/5}$ as in Figs 3c-5c. The maximum 2-m-small-scale buoyancy frequency '$N_{max}$' is indicated.

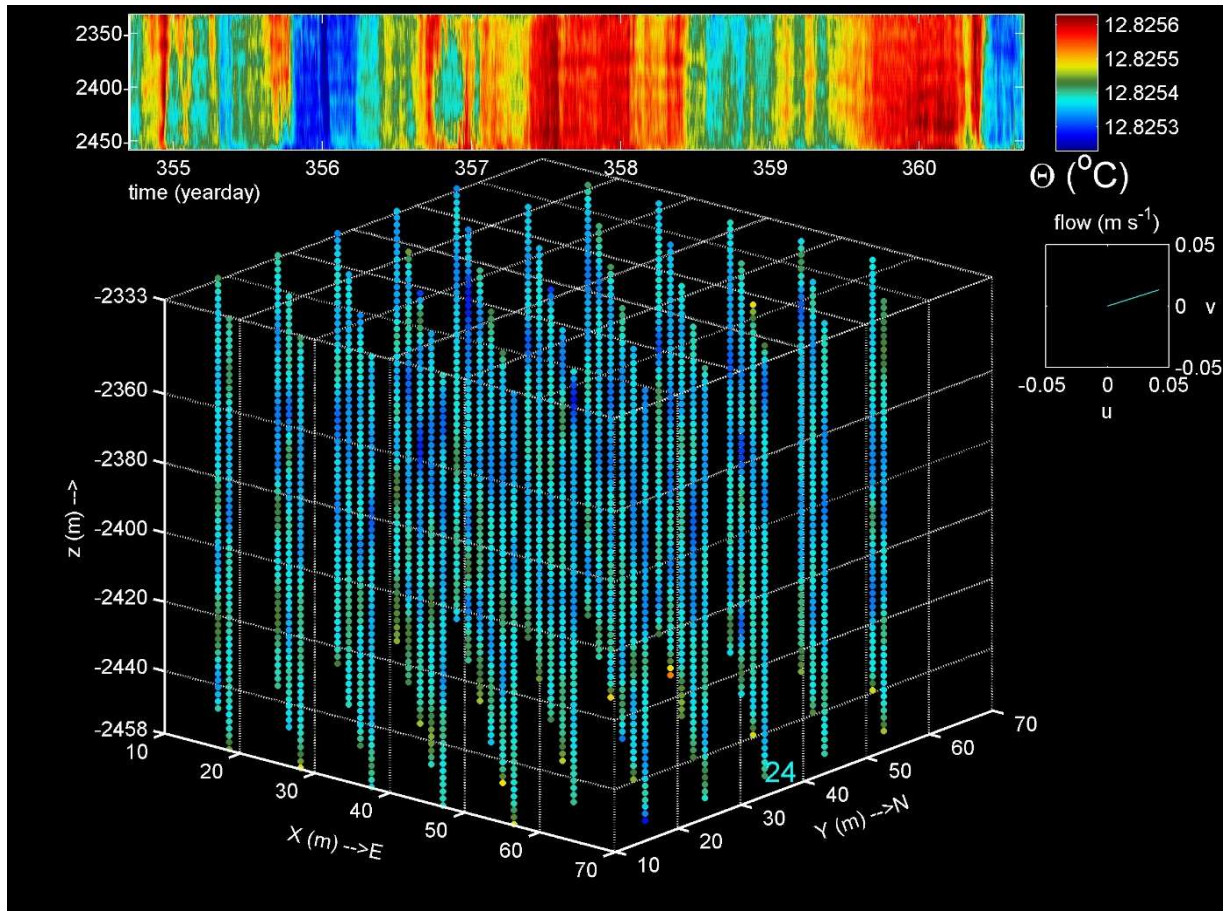

**Figure A1.** As Fig. 2 including movie, but for 6-day period during the second week of NH in Fig. 1.